\documentclass{cernyrep} 
\usepackage{texnames}
\usepackage[T1]{fontenc}
\usepackage[utf8]{inputenc}
\usepackage[bookmarks, colorlinks=true, linktoc=page, pdftex, linkcolor=blue, citecolor=blue, urlcolor=blue]{hyperref}
\begin{document}
\title{Chapter 5: Physics Opportunities with the FCC-hh Injectors}
 
\author{
B. Goddard\,$^{a1}$,\,
G. Isidori\,$^{a2}$,\,
F. Teubert\,$^{a1}$\,
(editors),\,
M. Bai\,$^{a3}$,\,
A. Ball\,$^{a1}$,\,
B. Batell\,$^{a4}$,\,
T. Bowcock\,$^{a5}$
G. Cavoto\,$^{a6}$,\,
A. Ceccucci\,$^{a1}$,\,
M. Chrzaszcz\,$^{a2,a7}$,\,
A. Golutvin\,$^{a1,a8}$,\,
W. Herr\,$^{a1}$,\,
J. Jowett\,$^{a1}$,\,
M. Moulson\,$^{a9}$,\,
T. Nakada\,$^{a10}$,\,
J. Rojo\,$^{a11}$,\,
Y. Semertzidis\,$^{a12, a13}$
\vspace*{2cm}
}

\institute{
$^{a1}$ CERN, CH-1211 Geneva, Switzerland \\
$^{a2}$ Physics Institute, University of Z\"urich, CH-8057
  Z\"urich, Switzerland \\
$^{a3}$ Forschungszentrum J\"ulich Institute for Nuclear Physics, 52425 J\"ulich, Germany \\
$^{a4}$ Pittsburgh Particle Physics, Astrophysics and Cosmology Center, Department of Physics and Astronomy, University of Pittsburgh, USA  \\
$^{a5}$ Departament of Physics, University of Liverpool, Liverpool, United Kingdom \\
$^{a6}$ Universita e INFN, Roma I, 00185 Roma, Italy  \\
$^{a7}$ H.Niewodniczanski Insitute of Nuclear Physics, PAN, Krakow, Poland \\
$^{a8}$ Imperial College, London, United Kingdom \\
$^{a9}$ Laboratori Nazionali di Frascati, 00044 Frascati, Italy\\
$^{a10}$ Ecole Polytechnique F\'ed\'erale de Lausanne (EPFL), Lausanne, Switzerland \\
$^{a11}$ Rudolf Peierls Centre for Theoretical Physics, 1 Kebble Road, University of Oxford, OXI 3NP Oxford, United Kingdom\\
$^{a12}$  Center for Axion and Precision Physics Research, Institute for Basic Science (IBS), Daejeon 34141, Republic of Korea\\
$^{a13}$ Department of Physics, Korea Advanced Institute of Science and Technology (KAIST), Daejeon 34141, Republic of Korea\\
}

\maketitle 

 
\begin{abstract}
In this chapter we explore a few examples of physics opportunities using the existing 
chain of accelerators at CERN, including potential upgrades. In this context the LHC 
ring is also considered as a part of the injector system. 
The objective is to find examples that constitute sensitive probes of New Physics that ideally cannot be done elsewhere or can be done significantly better at the CERN accelerator complex. Some of these physics opportunities may require a more flexible injector complex with additional functionality than that just needed to inject protons into the FCC-hh at the right energy, intensity and bunch structure. Therefore it is timely to discuss these options concurrently with the conceptual design of the FCC-hh injector system.
\end{abstract}
 
 \newpage

\section{Introduction}
\label{sec:Inj_intro}

The main problem we have in High Energy Physics (HEP) today is that we know there is physics 
beyond the Standard Model (SM), as indicated for example by the existence of Dark Matter, and by the fact that most probably New Physics (NP) is also needed to explain the origin of neutrinos masses, but we don't know at what energy scale(s) this NP appears. Conceptually, the SM does not include gravitational interactions, it has no explanation for the replication of 
quark and lepton flavours and suffers from problems related to the "unnatural" 
choice of its fundamental parameters. It may very well be that results from the 
LHC in the coming  years will give an unambigous indication of the relevant 
energy scale(s). Or it may be that FCC will provide this insight. But it could also be that the energy scale(s) are beyond what the LHC 
and even FCC can probe, which means that complementary approaches to increasing the available energy 
with the FCC-hh accelerator complex can be of enormous scientific interest.

\begin{figure*}
\centering
\resizebox{0.90\textwidth}{!}{%
  \includegraphics{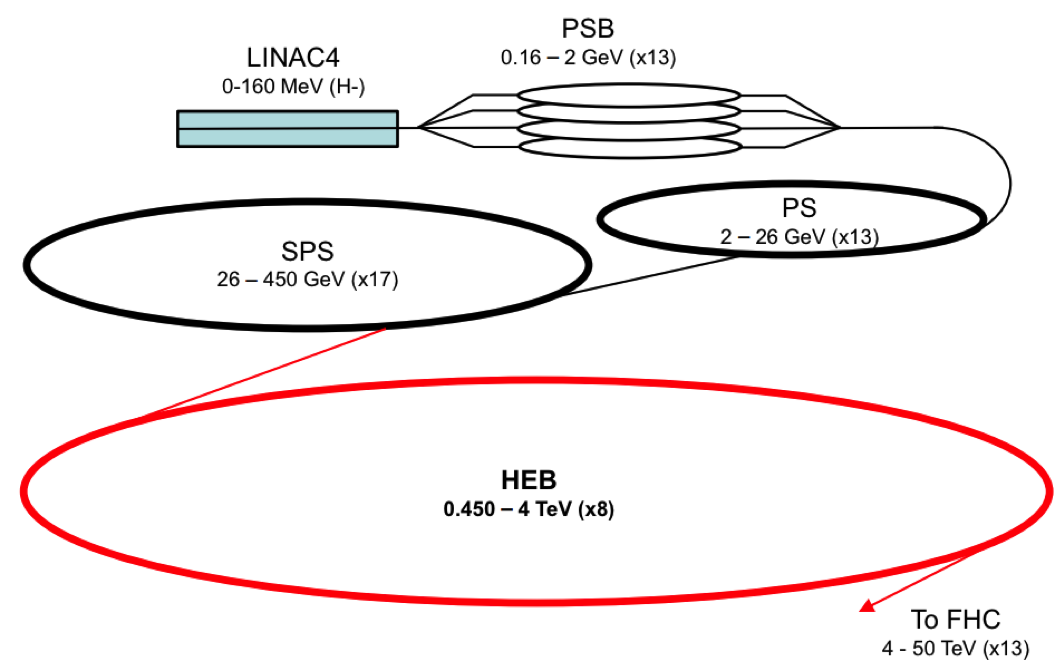}
}
\caption{Schematic view of the CERN accelerators system viewed as injectors of the future FCC-hh collider (FHC).}
\label{fig:1.1}
\end{figure*}

In this chapter we explore a few examples of physics opportunities using the existing 
chain of accelerators at CERN, including potential upgrades. In this context the LHC 
ring is also considered as a part of the injector system, becoming the High Energy Booster (HEB), see 
Fig.\ref{fig:1.1}. The objective is to find examples that constitute sensitive probes of NP that ideally cannot be done elsewhere or can be done significantly better at the CERN accelerator complex. Some of these physics opportunities may require a more flexible injector complex with additional functionality than that just needed to inject protons into the FCC-hh at the right energy, intensity and bunch structure. Therefore it is timely to discuss these options concurrently with the conceptual design of the FCC-hh injector system.

We will briefly discuss in the next sections the option to extract multi-TeV protons from HEB for tests of new detector concepts, the use of HEB for a very high luminosity experiment in collider mode when HEB is not injecting protons to the FCC, or several proposals to use high intensity beams of 400 GeV protons from the SPS for fixed target experiments. If polarized protons are available from the linear accelerator (LINAC4)  one could also envisage to extract 0.7~GeV polarized protons (the "magic" momentum) from the Proton Synchroton Booster (PSB) into a $\sim (40-120)$m radius storage ring to measure the proton Electrical Dipole Moment. In some of these examples, non trivial changes in the design of the FCC-hh injectors are required.

\newpage
\section{CERN hadron injector complex for FCC}
\label{sec:HI_1}

The FCC hadron collider will require an injector chain capable of filling both rings with around 10'000 bunches of a few TeV protons. 
Studying the FCC implementation at CERN naturally involves the reuse of the present tunnel infrastructure and hadron machines, up to and possibly including the LHC. With a nominal FCC-hh injection energy of 3.3~TeV~\cite{FCC_parameters}, the present LHC injector chains for protons and ions need to be complemented with a HEB, which will perform the acceleration from 450~GeV at exit from the SPS. Reuse of a suitably modified LHC is clearly a prime option for the HEB. 

\subsection{FCC-hh proton pre-injector chain}


The beam parameters required for FCC-hh \cite{FCC_parameters} are compatible with those delivered routinely to the present LHC machine at 450~GeV \cite{EVIAN_2015_LHC_performance}. The assumed FCC-hh proton injector chain is shown in Fig.\ref{fig:1.1}. 


Apart from the HEB, and aside any considerations of equipment lifetime, the HL-LHC injector chain could remain largely as it is for the FCC era. The parameters of the circular injectors in this chain are shown in Table~\ref{p_chain_parameters}, as expected after the upgrades for HL-LHC.

\begin{table}[hbt]
   \centering
   \caption{Parameters of proton accelerators in the LHC injector chain for LHC beams, after the LHC Injector Upgrades (LIU) for HL-LHC.}
   \begin{tabular}{lrcccc}
       \textbf{Parameter} 		& \textbf{Unit}	& \textbf{PSB} & \textbf{PS} & \textbf{SPS}\\
         Circumference 		&  m			& 157		& 628		& 6912  \\
	Extraction energy   		&  GeV		& 2.0			& 26			& 450  	 \\
         Cycle period    		&  s          		& 1.2			& 3.6			& 22.8   	 \\
         Max beam population     	& $10^{13}$   	& 1.2			& 2.0			& 7.7     	\\
         Beam power  			& kW 			& 3.2			& 23			& 240   		\\
   \end{tabular}
   \label{p_chain_parameters}
\end{table}

\subsection{FCC High Energy Booster options}

Three main options for the HEB are being studied for FCC-hh at CERN. The key parameters which will eventually decide between the different options are attainable FCC-hh injection energy, collider filling time, availability and, of course, capital and operating costs. 

While 3.3~TeV is the baseline FCC-hh injection energy for the study, optimisation is planned and a different eventual HEB energy is very likely. This parameter affects the performance of the collider in terms of beam physics and also in terms of magnetic field swing and harmonic control. Higher energies would reduce some constraints in the FCC-hh collider, while increasing the challenges for the HEB and its transfer lines. Lower energy injection will simplify the HEB and injector chain, at the cost of extra complexity, reduced performance limits and risk to the collider. 


Reusing the LHC as HEB offers a number of advantages over the alternative of building a completely new HEB, given the relatively limited changes needed to convert the existing, and very well-known, LHC machine into the final piece of the injector chain. Alternatively, a new accelerator based in the LHC tunnel using $\sim6~T$ fast ramping superconducting dipoles could be considered.

\begin{figure}[bht]
  \centering
  \includegraphics*[width=120mm]{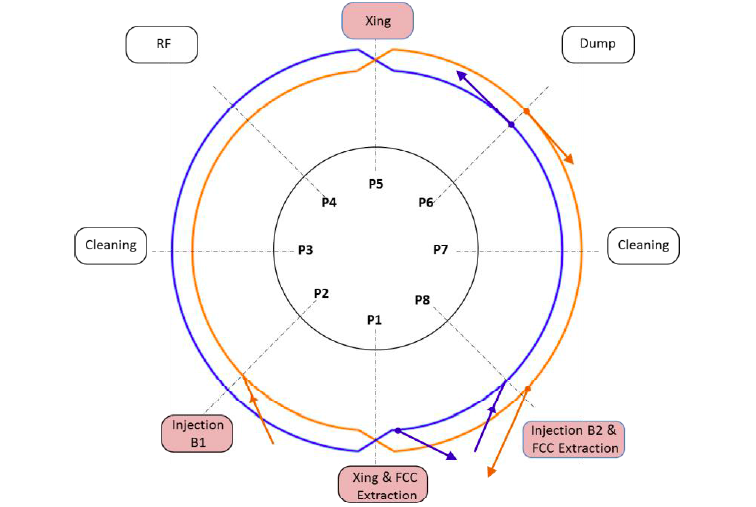} 
  \caption{Main functional layout of LHC as HEB as FCC-hh injector}
  \label{Fig_LHC_HEB}
 \end{figure}

For existing LHC re-use both rings are needed, to minimise FCC collider filling time. At least two crossings are needed to keep the rings the same length, with injection, collimation, RF and beam dump locations unchanged. The experiments and associated insertions in Points 1,2,5 and 8 would be decommissioned, and new extraction systems accommodated in P1 and P8 for transfer to FCC-hh. The new layout is shown schematically in Fig.~\ref{Fig_LHC_HEB}. The details of the layout and the main changes needed to the LHC are given in \cite{LHC_HEB_modifications}. In addition, the ramp rate would need to be increased by a factor of about five to fill the collider quickly enough, for which first feasibility considerations have been investigated~\cite{LHC_HEB_fast_ramp}.

The long straight section in LHC Point 5 which presently houses the CMS experiment would be free in this configuration, for the possible accommodation of an extraction system for Fixed Target beams, or for a high luminosity experimental interaction point (IP) further discussed in Section~\ref{sec:HI_3}.

A second option for the HEB is a compact Booster in the 7~km SPS tunnel. Limiting the dipole field to around 7~T (considered as a reasonable maximum for rapid-cycling at up to 1~T/s), the injection energy for the FCC collider would be 1.5~TeV. The accelerator would be single aperture with two extraction straights, a collimation straight, a beam dump straight, an injection straight and a RF straight, which with the present SPS tunnel six-fold symmetry does not leave a spare straight for other dedicated purposes like slow extraction. The study will need to determine whether an injection energy of 26~GeV is adequate, given the swing in energy of above 50. For this option, the simultaneous conception as a 2$\times$1.5~TeV collider with a high luminosity IP looks improbable, as this functionality would completely dominate the HEB design and cost.


The final option under consideration for the HEB is a 100~km accelerator in the FCC-hh tunnel, which would be relatively low-cost (at least per unit length) and fast cycling with iron-dominated magnets, but which would need to be powered by a superconducting drive cable to avoid prohibitive power consumption. This so-called super-ferric option would have a low magnetic field of about 1~T to reach 3.3~TeV (if the full machine arcs were filled with dipoles). Again, the conception of this HEB option as a collider seems improbable given the extra functionality required.


The parameters and main features of the SPS and the different HEB options are compared in Table~\ref{HEB_options}, including the potential beam power and estimated annual Proton on Target (p.o.t.) for an eventual Fixed Target operation. For the SPS, the two columns represent a well-analysed \cite{SHiP_p.o.t.} case of a short cycle (e.g. CNGS or with a very short 1~s spill) operating in parallel with a long cycle with a spill of around 10~s length. For the HEB options, the spill length is chosen to limit beam power on the target during the spill a rather arbitrary 5~MW, as this parameter may well be an important performance limitation at these very high primary beam energies and long spill lengths. For the HEB in the FCC tunnel, the ramp rate is limited by the RF system and not the ramping of the dipoles.


If all the available time that the injectors are not filling with protons the FCC-hh collider is used for other purposes (fixed target experiments, or other specific experiments located at the injectors), this will be about 80\% of the SPS and 60\% of the HEB operation.  Realistic factors for operational efficiency and transmission to the targets are included in the final p.o.t. estimates shown in Table~\ref{HEB_options}.

\begin{table}[hbt]
   \centering
   \caption{Parameters of SPS and HEB options for Fixed Target beams}
   \begin{tabular}{lrcccc}
       \textbf{Parameter} 		& \textbf{Unit}   & \textbf{SPS short/long} & \textbf{HEB@SPS} & \textbf{HEB@LHC} & \textbf{HEB@FCC} \\
         Extraction energy   		&  TeV	& 0.4		& 1.5	   	& 3.3		& 3.3		\\
         Dipole field		         	& T			& 1.8		& 6.7	   	& 3.9		& 1.0		\\
         Dipole ramp rate		& T/s			& 0.6		& 0.1		& 0.006	& 0.008	\\
        	Beam intensity		& $10^{13}$ p	& 4.5		& 6.3		& 30		& 120	\\
	Stored beam energy	& MJ			& 2.9		& 15		& 158		& 630	\\
         Min. Repetition period    	&  s          		& 7.2/16.8	& 135  	& 950		& 900	\\
         Beam power  			& kW 			& 400/170	&110	   	& 170		& 700	\\
       	Spill length          	  	& s 			& 1/10  	& 3		& 32		& 125	\\
         Peak extracted power 	& MW 			& 2.9/0.3	& 5.0		& 5.0		& 5.0		\\
       	FT hours per y	 	& h 			& 4000	& 3000	& 3000	& 3000	\\
       	Annual p.o.t. 	 		& $10^{19}$ p 	& 4.3/1.0	& 0.4		& 0.3		& 0.9		\\
   \end{tabular}
   \label{HEB_options}
\end{table}

An important caveat to note is that the annual p.o.t. quoted are realistic in terms of the cycling rate and machine intensity, but not necessarily as regards beam loss, activation and radioprotection constraints. These possible limitations are dicussed in Section~\ref{sec:HI_2}.

\section{Possibilities for a high-luminosity collider experiment in the HEB}
\label{sec:HI_3}

If the modified LHC is used as HEB, an interesting option to investigate is to have an experiment able to take proton-proton collisions at very high luminosities ($O(10^{35} \rm{Hz/cm}^2$)) with good trigger and acceptance for "low $\rm{P}_{\rm{T}}$" events (B, D, K and $\tau$ decays). ATLAS and CMS will have already cover most of the "high $\rm{P}_{\rm{T}}$" physics program at the LHC available energy by the end of the HL-LHC running at instantaneous luminosities of $5\times 10^{34} \rm{Hz/cm}^2$ for a total of $\sim 3 \rm{ab}^{-1}$ integrated luminosity. However, the LHCb current upgrade plan is to take proton-proton collisions up to instantaneous luminosities of $2\times 10^{33} \rm{Hz/cm}^2$ and integrate $\sim (0.05-0.1) \rm{ab}^{-1}$. It is clear that there is a window of opportunity to extend significantly this physics program using the LHC ring (both during the HL-LHC and the HEB eras). One could envisage to perform also this sort of experiment as an additional interaction point at the FCC ring at higher energies; however, the gain in cross section at those higher energies (a factor $\sim 5$)~\cite{FCCBBXS} needs to be balanced against the increase in occupancies and available luminosity. It could very well be that while an LHCb-like experiment at the FCC-hh collider can provide opportunities to study heavy flavour production in a region of phase space previously unexplored, an LHCb-like experiment at the HEB ring at very high luminosities can still be the best chance to study the B,D,K and $\tau$ decays with unprecedented precision.

In the next sections we consider the physics case for such an experiment at HEB, and then discuss possible scenarios in terms of accelerator performance. 

\subsection{Physics opportunities with a very high luminosity pp collider}
\label{XFLAVOR}
The indirect search for NP through precision measurements in flavour transitions is as strongly motivated as ever. It could very well be that in the coming years unambigous evidence of NP is seen in experiments looking at neutral flavour changing transitions of quarks, like LHCb, NA62, KOTO or Belle-II, and/or experiments specialized on neutral flavour changing transitions of leptons, like MEG-II, Mu2e, COMET-II, neutrino oscillation experiments, etc... In such a situation, the case to improve on the precision will be obvious. Furthermore, even without evidence for NP, an LHCb-like experiment at the HEB ring able to take data at a rate of two orders of magnitude above the current LHCb upgrade design will provide a unique opportunity to constrain viable models of a flavour theory.

Very rare decays which will be limited by the statistics available are a clear case. For instance, the decay $B_{\rm d} \rightarrow \mu^+ \mu^-$ will be measured with a precision not better than $30\%$ by ATLAS, CMS and LHCb due to the limited number of candidates, O(100),  in data by the end of the HL-LHC era assuming similar sensitivity as achieved in RUN-I~\cite{BMUMU_LHC}. If an LHCb like experiment in the HEB ring could increase this to O($10^4$) candidates, it could allow to determine the ratio: BR($B_{\rm d} \rightarrow \mu^+ \mu^-$ )/BR($B_{\rm s} \rightarrow \mu^+ \mu^-$ ) to a percent precision, allowing for a very stringent test of flavour models. Another excellent example is the search for lepton flavour decays in $\tau^{\pm} \rightarrow \mu^+ \mu^- \mu^{\pm}$. The enormous charm and beauty production at proton-proton colliders and subsequently of $\tau$ leptons from their decays has opened a new window of opportunity increasing the $\tau$ production rate by five order of magnitude w.r.t. the $e^+ e^-$ B-factories. Indeed, LHCb expects to reach sensitivities of O($10^{-9}$) with their upgrade~\cite{TAU3MU_LHCB}, comparable with the expected sensitivity from Belle-II~\cite{TAU3MU_BELLE}. The five orders of magnitude increase in production rate at LHCb compensates for the lower efficiency in a proton collider O($10^{-2}$) and lower integrated luminosity (0.05 $\rm {ab}^{-1}$ at LHCb vs 50 $\rm {ab}^{-1}$ at Belle-II). An LHCb-like experiment at the HEB ring could reach sensitivities of O($10^{-10}$) allowing for a strong test complementary to the searches for lepton flavour violation in muon decays. Similar sensitivities seem to be feasible in a dedicated fixed target experiment at the SPS, see Section~\ref{DARK}.

Moreover, if the detector developments in the coming years proves feasible to have tracking devices that provides also accurate timing information (O(10 ps)), it should be possible to identify the production vertex in such a pileup environment (O(1000))~\cite{INFN_WHATNEXT}. In such a case a time dependent analysis of CP asymmetries in $B_{\rm s} \rightarrow J/\Psi \mu^+ \mu^-$ decays should, for example, allow to improve on the determination of the phase in the $B_{\rm{s}}$ mixing. Indeed, ATLAS, CMS and LHCb expect to reach sensitivities of $\sim 5~\rm{mrad}$ in $\phi_{\rm s}$, i.e. the phase of the $V_{\rm{ts}}$ CKM coupling, by the end of the HL-LHC era extrapolating the current sensitivities~\cite{PHIS_LHCB, PHIS_CMS, PHIS_ATLAS}. However, the indirect determination of $\phi_{\rm {s}}$ is already known with a precision better than 2 mrad from "tree level" measurements~\cite{CKMFITTER}, i.e. measurements mostly not affected by NP in the loops. An LHCb-like experiment at the HEB ring could reach a precision better than 1 mrad allowing for a precise comparison between "tree measurements" and "loop measurements" to disentangle possible NP contributing to the $B_{\rm{s}}$ mixing with similar precision as the B-factories have done in the past for the $B_{\rm{d}}$ system.

\subsection{Possible high luminosity insertion design in the different HEB lattices}

At this preliminary conceptual stage, a dedicated high-luminosity experiment in the 7~km SPS or 100~km FCC-hh tunnels is not considered feasible, since building the HEB from scratch as a collider would dominate the design requirements, and increase significantly the cost and complexity of what should be a minor part of the FCC project. The discussion here is therefore restricted to the case of the re-use of LHC as HEB. 

The beam parameters assumed for initial performance evaluation are those of HL-LHC: 3.3~TeV, 2.2$\times10^{11}$ p/bunch at 25~ns spacing, and 2.5~$\mu$m emittance. A 15~cm $\beta^{*}$ with a 12~$\sigma$ separation requires a crossing angle of about 0.86~mrad, and hence a crab-crossing system to prevent a punitive geometric reduction factor. Such a crab crossing system is part of the HL-LHC baseline, and so can be assumed as accessible for the FCC era. Under these conditions the instantaneous luminosity exceeds $10^{35} \rm{Hz/cm^2}$. For a collider operating for $60\%$ of the time, over a 200 day run the integrated luminosity would be approximately $250~\rm{fb}^{-1}$, assuming a H\"ubner factor of 0.2 \cite{Hubner}. Operating the HEB collider at 6.5~TeV would effectively double this, to $2.2\times10^{35}$cm$^{-2}$s$^{-1}$ instantaneous luminosity and $500~$\rm{fb}$^{-1}$ per year. These parameters do not appear impossible in terms of maximum beta, triplet aperture, tune shift, beam brightness and stored energy.

To reach 1~ab$^{-1}$ per year at 6.5~TeV (or 0.5~ab${-1}$/y at 3.3~TeV) is more challenging and would require more ambitious parameters, for instance operating at 2.4$\times10^{11}$p/bunch with 2.0~$\mu$m emittance (corresponding to 33\% higher brightness in collisison than the HL-LHC baseline) and 10~cm $\beta^{*}$. Here the feasibility depends on whether the assumed performance of the injector chain for HL-LHC can be improved upon by such a large factor, as well as the feasibility of operating with a 10~cm $\beta^{*}$ optics.

\section{Performance reach of Fixed Target beams extracted from the FCC injectors}
\label{sec:HI_2}

The motivation for the provision of Fixed Target (FT) beams is for specific physics experiments where the kinematics and experimental characteristics suit best this type of experiment, and also for detector test beams, where fluxes of high energy particles are needed to characterise and develop detector concepts and sub-assemblies. In addition, materials testing and radiation resistance may also require specific test beam characteristics, as exemplified by the HiRadMat and CHARM facilities in operation today at CERN \cite{HiRadMat, CHARM}.

The limitations of the average power on the target are often linked to beam losses in the extraction region and around the accelerator, which are much higher for slow extraction than for fast extraction. The detector technology and performance therefore has a direct impact on the effective beam power and protons on target (p.o.t.) that can be achieved. 


\subsection{Extraction types and limitations for Fixed Target beams}

In general, detector technology means that FT physics experiments and test beams require relatively long-duration, constant flux of particles, requiring slow extraction methods \cite{slow_extraction}. The key factors are flux, stability and duty cycle. The SPS slow spill can provide 400~GeV protons over about 10~s, with a total stored beam energy of up to 3~MJ approximately every 16~s. 

The exceptions are typically those physics experiments where strong pulsed focussing elements are needed to produce the required secondary beam characteristics and where detector occupancy is less of an issue, or materials testing, both of which require much shorter extracted beam time structure and can be served with fast single-turn extraction. Examples at CERN are the 450~GeV HiRadMat materials test beam, which contains about 3~MJ of energy in a 8~${\mu}s$ spill, and the 400~GeV CNGS beam, which contained two spills of 10.5~${\mu}s$ spaced at 50~ms intervals each containing about 1~MJ of energy every 6 seconds. In these domains target design constraints and target area irradiations considerations are also very important.


In the SPS the slow extraction system already fully occupies one of the six straight sections, using 90~m of space. At even higher energies the design of an extraction system becomes increasingly problematic. Without new technical developments a conventional slow extraction system will be difficult for 3.3~TeV beam energies, especially for an accelerator with a superconducting main magnet system, given the low energy deposition limits in superconducting magnet coils. A significant study program is needed to investigate whether slow extraction could be compatible with a superconducing HEB, and to investigate the system design. There are aleady some promising technical directions: for instance a bent crystal could possibly replace an electrostatic septum in the slow extraction channel, to provide a much more compact and radiation resistant system. The use of bent crystals to provide strong deflections to high energy protons has been demonstrated experimentally, and it is a possible new technology route to a slow extraction system at multi-TeV energies \cite{crystal_02}. Investigations and theoretical studies are ongoing in SPS and LHC.


The achievable p.o.t. is limited by the beam intensity, the beam energy, acceptable beam loss rates and also by the accelerator cycling rate, which for small accelerators is inherently faster than for large machines. Beam losses and activation are major design issues for the extraction systems, affecting machine operation, through the limitations on personnel doses, shutdown lengths, access restrictions, the need for remote handling and cooldown-times. Ultimately these aspects often limit the achievable p.o.t., rather than the accelerator peak performance.


Increasing the protons available from SPS above the ballpark reach of 5$\times10^{19}$ p.o.t./y presently demonstrated with fast extraction will rely on a combination of factors in the CERN complex. One is improved beam loss control associated with the production of the FT beam in the PS, for which the improvements in beam brightness from the HL-LHC related upgrades combined with a Multi-Turn Extraction approach \cite{MTE_01} is a possible solution. A target of 7$\times10^{13}$~p per SPS cycle has been discussed as a realistic target in the past \cite{PAF_report}, which seems reasonable as the HL-LHC beams will be closer to 8$\times10^{13}$~p per (longer) SPS cycle. The beam losses in the SPS during injection, capture and acceleration also needs to be tightly controlled, but these will be overshadowed by the losses associated with the slow extraction process itself. Here a significant improvement is needed to keep activation levels reasonable. This could be via new approaches to the extraction, new technology for extraction equipment, remote handling or new dectector technology allowing fast extraction spills. Finally, the primary beam targets will also eventually limit the annual p.o.t., either through thermomechanical stress effects or average beam power, or from activation in the target areas.

If extraction losses and activation contraints can be overcome, increasing the SPS FT beam intensity to 7$\times10^{13}$~p per cycle could allow to reach the region of 8$\times10^{19}$~p.o.t./year.

\subsection{Future opportunities using fixed-target beams at FCC-hh pre-injectors}

\subsubsection{Kaon Physics}

One of the strongest constraints on the possible size of NP contributions comes from Kaon physics, in particular the precise measurement of the mass difference ($\Delta m_K = m(K_L)-m(K_S)$) and the CP-violating quantities $\epsilon_K$ and $\epsilon'$. This is because the SM suppression factors are smaller in the Kaon sector, since the u and c-quark contributions to FCNC processes are very strongly suppressed by the Glashow-Ilioupoulos-Maiani (GIM) mechanism, while that of the t-quark is strongly suppressed by the Cabbibo-Kobayashi-Maskawa (CKM) matrix elements. There is great interest in decays with a neutrino pair in the final state like $K^+\rightarrow \pi^+ \nu \nu$ and $K_L^0 \rightarrow \pi^0 \nu \nu$ as they are determined by short distance physics. In these cases, there is a single operator that determines the decay rates within the SM and in most NP scenarios.

The NA62 experiment~\cite{NA62} at CERN has the potential to measure the $BR(K^+ \rightarrow \pi^+ \nu \nu)$ with at least a $10\%$ precision. With an expected signal acceptance of $\sim 10\%$ and S/B>4.5, the experiment requires $\sim10^{13}$ K decays to achieve such goal. The CERN SPS provides $3 \times 10^{12}$ 400 GeV protons on target per 16.8 second spill, which produces a very high intensity K beam, resulting in 5 million Hz K decays in a 60m long vacuum chamber. The sample available to the NA62 experimenters corresponds to $\sim4.5\times10^{12}$ K decays whose flight path is in their acceptance per year ($\sim10^7$sec). They expect to see $\sim45$ SM signal candidates per year with <10 background events~\cite{NA62}. 

The KOTO experiment~\cite{KOTO} at KEK has the potential to reach a first observation of the decay $K_L^0 \rightarrow \pi^0 \nu \nu$ at the level of the SM prediction, and has the goal to upgrade the facility such that it allows for a $\sim10\%$ measurement of the branching fraction. The J-PARC accelerator is designed to provide $2\times10^{14}$ 30 GeV protons on target every three seconds. Moreover, the neutral K beam is highly collimated (pencil beam) so that the reconstructed $\pi^0$ momentum component transverse to the beam direction can be used as a constraint. 

Preliminary studies in the context of a PRIN grant in Italy, (KLEVER, PRIN call 2010-11), have concluded that it should be possible to reach a sensitivity similar to the future KOTO sensitivity if a high intensity 400~GeV proton beam ($10^{13}$ protons on target per 16.8 s spill, corresponding to $\sim~10^{19}$~p.o.t./year) is available from the SPS. To improve significantly higher proton intensities are required and significant detector research and development.  This intensity will be beyond what is achieved now at the T10 facility in the north area at CERN, due to various constraints and beam-sharing, but could be possible if this facility is upgraded, or if a generic new north area high intensity facility is build to serve other experimental proposals like SHiP~\cite{SHIP_PHYSICS} or other experiments searching the Dark Sector.

\subsubsection{Dark Sector}
\label{DARK}
The fact that no clear evidence for NP has been observed so far from precision measurements below the EW energy scale has to be due to the effects of NP being highly suppressed. This can be because the mass scale of NP particles is sufficiently larger than the EW energy scale, and/or because the couplings are small or some new symmetry acts such that the effects are cancelled. The Dark Sector generally refers to the possibility that NP particles with masses below the EW energy scale have not been detected because their interactions with SM particles are highly suppressed. The Dark Sector is usually classified in terms of the operators which mediate their interaction with SM particles. These interactions are the "Portals" to the Dark Sector.

If these NP particles are below the D/B mesons mass they can be produced with a Fixed Target experiment at the SPS due to the large D and B meson production cross-section at the SPS energy. This is the basic concept for the SHiP proposal, aiming to collect data at the largest possible intensity of the SPS proton beam. Detailed studies of the neutrino portal, i.e. searches for heavy neutral leptons (HNL) particles have been performed assuming $\sim4 \times 10^{13}$ protons per spill~\cite{SHIP_PHYSICS}. The beam and particle backgrounds are suppressed by adding a particle filter downstream of the target, allowing only these Dark Sector particles (plus some residual muons and SM neutrinos) to reach a long decay tunnel equiped with detectors. For example, SHiP sensitivities studies~\cite{SHIP_PHYSICS} shows an expected improvement by two order of magnitude of the limits on the couplings in the HNL mass region (1-2) GeV. In the same mass region, a running experiment like NA62 could potentially improve by one order of magnitude the existing limits, but cannot reach the expected sensitivity of the SHiP design. For larger masses (but still below the Z/W mass) other high energy facilities (like LHC or FCC-ee) are probably more appropriate for this kind of search.

In the same SHiP study~\cite{SHIP_PHYSICS} a modified detector setup (including a specific target design) shows the potential to search for lepton flavour decays in $\tau^{\pm} \rightarrow \mu^+ \mu^- \mu^{\pm}$ in a specific designed fixed target experiment, reaching sensitivities of O($10^{-10}$) similar than in Section~\ref{XFLAVOR}.

\section{Polarized protons}
\label{sec:HI_4}


The potential of polarized hadron-hadron collisions to characterize NP were already discussed in~\cite{RHIC_POL1, RHIC_POL2, RHIC_POL3} in the context of the RHIC physics program. More recently, some authors have discussed the benefits of having polarized protons in the FCC ring to disentangle the couplings of NP particles to the different SM quarks, and as a tool to significantly reduce SM backgrounds~\cite{FCC_POL}. 
In the context of this chapter, having polarized protons in the injector chain will also open new physics opportunities. One clear example is the measurement of the electric dipole moment (EDM) of nucleons. Within the SM, the EDM of a nucleon is expected to be below $10^{-31} \rm{e}\times \rm{cm}$~\cite{EDM_SM}. The current limits using ultracold neutrons in a bottle reaches a sensitivity of  $2.9\times 10^{-26} \rm{e}\times \rm{cm}$ at 90$\%$C.L.~\cite{NEUTRON_EDM_LIMIT}. Future neutron facilities in Europe and elsewhere should be able to improve significantly this sensitivity. In the case of protons the sensitivity is about an order of magnitude worse and indirectly inferred from $^{199}\rm{Hg}$~\cite{PROTON_EDM_LIMIT}. An observation of a non-zero nucleon EDM in the near future will be a clear indication of NP and of new sources of CP violation in strong interactions until the sensitivity reaches the level of the SM predictions. 

\subsection{EDM storage rings}

Several groups in the USA and in Europe~\cite{EDM_PROPOSALS} have been developping plans for a storage ring with spin coherence times of about $10^3$ seconds and electrical gradients of 4.5 MV/m, being fed with $10^{11}$ protons per cycle with $80\%$ polarization at a "magic" momentum of 0.7~GeV/c reaching sensitivities of O($10^{-29} \rm{e}\times \rm{cm}$). 

For protons, an all-electric storage ring is possible, at 0.7~GeV/c where  the spin and momentum vectors precess at the same rate in any electric field. The radial E-field acts on the proton EDM and can cause a measurable vertical spin precession.

Such facilities are under extensive study, and will rely on the provision of high intensity polarized proton beams. The low energy polarized protons at the start of the FCC injector chain could be used for such EDM rings.  One could extract polarized protons from the PSB and inject them into a relatively small storage ring with a radius on the range between 40~m and 120~m. To improve on sensitivity it is foreseable to be able to improve the spin coherence time by an order of magnitude ($10^4$ seconds) and use new techniques like the stochastic cooling-thermal mixing and higher proton beam intensities. If in addition, a reliable electrical gradient of about 15~MV/m with negligible dark currents is achieved, then the sensitivity of such a ring will reach the SM expectations for proton EDMs. Proposals for deuteron EDM storage rings have also been made~\cite{EDM_DEUTERON}.

In the following section we discuss briefly what sort of performance could be expected from the injector system as well as the requirements to the injector system to maintain the polarization of protons to be injected into the FCC.

\subsection{Polarized beams for FCC in light of RHIC experience}

RHIC has sucessfully and routinely accelerated and collided polarized protons up to about 250~GeV, using a dual spin-rotator (Siberian snake) setup \cite{RHIC_polarisation}. An average of 55\% store polarization was achieved. The difficulty is to overcome depolarising resonances - the imperfection resonances are separated by only 523~MeV, so about 50 of these resonances would be crossed in the PS, around 1000 in the SPS and over 5'000 in the 3.3~TeV HEB. In addition the resonances get stronger with higher energy. 

For the FCC injector chain, a new polarized proton source would be needed, together with extensive changes in all of the circular pre-injectors, where the depolarising resonances can either be compensated (in the low energy PSB) or overcome with the use of Siberian snakes. The integration of the required snakes in the existing PS \cite{CERN_PS_DL_76-9} and, to a lesser extent, the SPS is likely to be problematic. The snake is a helical dipole, several meters long, which rotates the vertical polarisation by 180~degrees, making the spin-tune a half-integer and energy independent and avoiding the imperfection or resonance conditions. The spin rotation of the snake has to be much larger than the total spin rotation from the resonances, which means many snakes are needed around the larger rings. The preservation of polarization in the HEB and FCC collider itself would be uncharted territory, although initial considerations for LHC have proposed 16 snakes with 2 per arc \cite{Roesser_LHC_polarisation}. A final complication is that an extreme control of residual orbit error is needed - to around 10$\rm\mu$m for LHC (presently about 200$\rm\mu$m). 

Altogether providing and colliding polarized proton beams for the FCC-hh collider appears to be a substantial challenge, and one which might significantly affect the design of the HEB and collider itself. However, the spectacular success of RHIC has demonstrated that providing and colliding polarized protons up to energies of several hundred GeV is perfectly feasible. For beams below the GeV range for an EDM facility, the similarity of the RHIC injector chain to the FCC pre-injectors \cite{MB_Polarized_protons_at_AGS} gives confidence that polarized protons or deuterons could be accelerated to the energies needed in a version of the PSB. Polarization levels of $\sim$80\% could be expected, for single bunch intensities of at least a few $10^{11}$p. 


\section{Summary}
\label{sec:HI_5}

Full exploitation of CERN's infrastructure in the FCC era would make best use of the proton injector chain during the time it is not filling the collider. These accelerators could be used to deliver a variety of beams to different facilities with the potential for unique physics reach. In addition, there is a clear case for high energy test beams to be fast- and slow-extracted from the HEB, for detector developments and also for materials and structural robustness characterisation of accelerator subsystems. 

A preliminary and certainly incomplete examination of the possibilities already gives an interesting list of possible physics opportunities, as outlined above:

\begin{itemize}  
	\item Reaching sensitivity down to the SM predictions on EDMs for nucleons using polarized protons in a dedicated 0.7~GeV storage ring; 
	\item Precision search for flavour-changing transitions through B,D,K and $\tau$ decays, in a dedicated HEB high-luminosity collider experiment;
	\item Search for BSM dark sector particles in a 400~GeV high intensity proton beam dump experiment from the SPS, as for example HNLs;
	\item Improve on sensitivity for $K_L^0 \rightarrow \pi^0 \nu \nu$ decay branching ratio using a 400~GeV high intensity proton beam from the SPS.
\end{itemize}

Most of the potential physics experiments require large fluxes of high energy protons on target. There is a strong coupling between the detector technology and the expected accelerator performance limits, through the losses and activation which result from slow extraction, imposed by the detector constraints. Increasing the annual p.o.t. from the SPS beyond the $5\times 10^{19}$ presently achievable with fast extraction will be an important challenge, as will be approaching this number for slow extracted beams. Overall, a number of specific aspects can be highlighted for possible study directions to determine the performance reach for the different applications, including:

\begin{itemize}  
	\item Preservation of polarization through the injector chain, and spin dynamics in a proton and/or deuteron EDM storage ring;
	\item High field (at least 15~MV/m) electrostatic bending elements;
	\item Proton spill structure and detector occupancy/pile-up limitations;	
	\item Beam loss reduction and activation mitigation for high intensity slow extraction, beam transport and target/experimental zones; 
	\item Beam dilution, target materials, layouts, robustness and handling;
	\item Potential for crystal extraction for high intensity, high energy protons;
	\item High luminosity experimental IP design for LHC as FCC HEB.
\end{itemize}

\newpage
\bibliographystyle{report}
\bibliography{report}

\end{document}